\documentclass[amsmath,preprintnumbers,superscriptaddress,nofootinbib,aps,prd,10pt,a4paper]{revtex4}
\usepackage[T1]{fontenc}
\usepackage[utf8]{inputenc}
\usepackage{array}
\usepackage{multirow}
\usepackage{amsmath}
\usepackage{amssymb}
\usepackage{graphicx}
\usepackage{makecell}

\usepackage{ulem}  % for delete
\usepackage{xcolor}  % for colored text

\makeatletter

\@ifundefined{textcolor}{}
{%
 \definecolor{BLACK}{gray}{0}
 \definecolor{WHITE}{gray}{1}
 \definecolor{RED}{rgb}{1,0,0}
 \definecolor{GREEN}{rgb}{0,1,0}
 \definecolor{BLUE}{rgb}{0,0,1}
 \definecolor{CYAN}{cmyk}{1,0,0,0}
 \definecolor{MAGENTA}{cmyk}{0,1,0,0}
 \definecolor{YELLOW}{cmyk}{0,0,1,0}
}

\usepackage{amsfonts}
\usepackage{array}
\usepackage{multirow}
\usepackage{latexsym}
\usepackage{epsfig}
\usepackage{float}
\usepackage{dsfont}

\usepackage{ragged2e}
\justifying\let\raggedright\justifying
\newcommand{\be}{\begin{equation}}
\newcommand{\ee}{\end{equation}}

\RequirePackage{doi}

\makeatother

\begin{document}
\title{Secondary Gravitational Waves in Non-local Starobinsky inflation}

\author{Andrea Addazi}
\affiliation{Center for Theoretical Physics, College of Physics, Sichuan University,
Chengdu, 610064, PR China}
\affiliation{INFN, Laboratori Nazionali di Frascati, Via E. Fermi 54, I-00044 Roma,
Italy}
\author{Alexey S. Koshelev}
\affiliation{School of Physical Science and Technology, ShanghaiTech University, 201210 Shanghai, China}
\affiliation{Departamento de Fısica, Centro de Matematica e Aplicacoes (CMA-UBI), Universidade da Beira Interior, 6200 Covilh\~{a}, Portugal}

\author{Shi Pi}
\affiliation{CAS Key Laboratory of Theoretical Physics, Institute of Theoretical Physics,
	Chinese Academy of Sciences, Beijing 100190, China}
\affiliation{Center for High Energy Physics, Peking University,  Beijing 100871, China}
\affiliation{Kavli Institute for the Physics and Mathematics of the Universe (WPI), The University of Tokyo, Kashiwa, Chiba 277-8583, Japan}

\author{Anna Tokareva}
\affiliation{School of Fundamental Physics and Mathematical Sciences, Hangzhou Institute for Advanced Study, UCAS,
Hangzhou 310024, China}
\affiliation{International Centre for Theoretical Physics Asia-Pacific, Beijing/Hangzhou, China}
\affiliation{Theoretical Physics, Blackett Laboratory, Imperial College London, SW7 2AZ London, U.K.}

\begin{abstract}
We show how infinite derivative modifications of gravity 
impact on the stochastic background of Gravitational Waves from early Universe. The generic property of the ghost-free theory fixed on Minkowski space-time is the emergence of an infinite number of complex mass states when other classical backgrounds are considered. These additional states are shown to enhance the power spectrum of scalar perturbations generated during inflation. 
Current and future space-based and terrestrial interferometers offer indirect testing methods for the infinite derivative gravity action, enabling the exploration of new parameter spaces.  In particular, we identify unconventional blue-tilted Gravitational Wave spectra, presenting a novel approach for testing infinite derivative quantum gravity in the future.

\end{abstract}
\maketitle
\tableofcontents{}

\section{Introduction}

As it is well known, Einstein's General Relativity (GR) is a non-renormalizable theory \cite{Wald:1984rg}. The idea, developed later on by Stelle \cite{Stelle:1976gc}, to modify GR to a renormalizable theory by adding curvature squared terms, is plagued by ghosts. This is not surprising: higher derivatives are known to generate ghost-like degrees of freedom \cite{Ostrogradsky:1850fid}. There is, however, a possible way out of this problem by
extending Stelle's proposal to an infinite derivative or non-local model. This was suggested and realized decades ago \cite{Kuzmin:1989sp,Tomboulis:1997gg}, and elaborated later on in \cite{Modesto:2011kw,Modesto:2014lga,Modesto:2015ozb}, by the insertion of infinite derivative operators or form factors in the quadratic gravity action. 
The key observation here is power-counting renormalizability, which applies to form factors with constant coefficients in front of derivative powers \cite{Koshelev:2017ebj}.

Several efforts were recently made in this direction but a lot of puzzles remain unanswered yet.
One of the most challenging issues is the need for fine-tuning of form factors to develop a theory without introducing any new degrees of freedom (d.o.f.). A model that incorporates an exponential of an entire function of the d’Alembertian operator is a significant step toward solving this problem \cite{Biswas:2011ar,Biswas:2016egy}. However, such a construction is somewhat ``fragile'': it does not survive when shifting to another vacuum state in a non-local field theory or gravitational metric background. 
Setting conditions to remove extra degrees of freedom in a specific background does not ensure their elimination in other backgrounds. Thus, as a rule, infinite derivative field theory or gravity will include extra d.o.f. in case of the existence of multiple background solutions. In general, such extra excitations will have complex masses squared, which are not usually considered in conventional field theories. In what follows we will dub them background-induced states (BIS).

It is worth noting that these surprising features are not limited to infinite derivative models but are also observed in some finite higher derivative models. For example, one could consider a sixth-order gravity theory, where the Lagrangian includes a kinetic term with six derivatives, or equivalently, cubic polynomials of the d'Alembertian operator. In fact, a cubic polynomial term may already correspond to one real and two complex conjugate roots, and the latter will appear as complex squared masses in the propagator. Under the assumption that our Lagrangian contains only real parameters, if complex masses are generated, they will always be in complex conjugate pairs. 

There are different ideas on how to deal with BIS. One of the proposals is to impose fine-tuned initial conditions to obtain certain cancellations and disappearance of such states from the spectrum \cite{Anselmi:2017yux}. In this case,  complex mass particles are effectively not included as on-shell asymptotic states but considered only as off-shell virtual fields. In addition, there are several arguments that such complex mass excitations could break causality and unitarity if not satisfying precise criteria on the momentum dependence of the graviton propagator \cite{Platania:2022gtt}. However, as mentioned above, in infinite derivative gravity, BIS can be elided only around one selected background. 
In fact, it seems to be impossible to tune the theory in such a way that there are no extra excitations around all possible backgrounds. One potential solution is to introduce background-dependent form factors, although this approach has several downsides \cite{Koshelev:2022olc}. Indeed, renormalizability demands at least a higher level of fine-tuning which can easily be spoiled by new interaction vertices. Additionally, 
 the tuning of background dependence can be straightforwardly imposed only on top of the maximally symmetric spacetimes;
this means that the problem will reappear in the case of any small deviations from these conditions.
Alternatively, as suggested in Ref.\cite{Tokareva:2024sct},
one can accept the existence of BIS and study their dynamical effects. This discloses a ``Pandora's box'' of issues concerning the implications of these states in the early Universe. 

In the present paper, 
as a first step in exploring BIS in cosmology, we analyze an infinite-derivative model of inflation. It is clear that, regardless of the specific model, there are at least two background configurations to consider: the Minkowski vacuum and the inflationary state. Given these considerations, unless special techniques or modifications are applied, infinite derivative form factors can only ensure a ghost-free spectrum around one of the background solutions. Indeed, Minkowski vacuum should have no extra pathological excitations. Indeed, the Minkowski vacuum should be free of additional pathological excitations. This logically implies that, in general, the inflationary stage will include new states with complex masses squared, i.e., the BIS.

To clarify the concept, let us consider the simplified case
of
the harmonic oscillator in a flat space-time with a complex mass.
As it is clear, this system is unstable, having at least one exponentially growing mode. Nevertheless,
this analysis is not straightforward in case of curved space-time.
For instance, one can find that, in a de Sitter space-time, complex mass states can be classically stable if the condition
\begin{equation}
\label{StabilityCondition}
\text{Im}(m^2)^2<9H^2 \text{Re}(m^2)
\end{equation}
is satisfied \cite{Koshelev:2020fok}
where $H$ is the Hubble rate.
Such a bound 
is crucial in Cosmology as
a ``razor criterion'' 
for separating classically stable and unstable configurations. 

Stable BIS will have rapidly decaying behavior without any impact on the observables in the Cosmic Microwave Background (CMB). However, unstable ones will absolutely destroy CMB predictions and jeopardize the inflation dynamics itself. We conclude from here that only masses that are inside the stability region can be safely considered. 
In a conservative and minimal approach, we will not consider cases altering CMB as the latter is measured with great accuracy. 
%, and we will assume a simple inflationary model is absolutely enough. Any additional ingredients have to be extremely fine-tuned.
%Therefore, we expect that BIS-s can be relevant only for effects that do not interfere with CMB measurements.
It is important to stress here that values of BIS masses are determined by the higher derivative form factors. Nevertheless, 
the behavior of these states is always governed by the Klein-Gordon equation with a complex mass squared in an inflationary (nearly de Sitter) background.

In cosmological perturbation theory, the power spectrum \(P(k)\) quantifies the distribution of perturbation amplitudes over different scales (or equivalently, wave numbers \(k\)). When \(P(k)\) is enhanced at large \(k\), it indicates that smaller scales (larger wave numbers) have larger perturbation amplitudes. This scenario can arise from various inflationary models where specific mechanisms amplify the scalar perturbations during inflation. 
%{\color{red} (SP: Previously \cite{Brandenberger:2012aj} was cited but it was about trans-Planckian physics on large scales, \textit{i.e.} small $k$. To enhance $\mathcal{P}_\zeta(k)$ on small scales, we need some special dynamics.)}
Scalar perturbations are the sources of density fluctuations, which can also source second-order tensor perturbations - gravitational waves - through nonlinear interactions \cite{Matarrese:1992rp,Matarrese:1993zf,Matarrese:1997ay,Noh:2004bc,Carbone:2004iv,Nakamura:2004rm,Ananda:2006af,Osano:2006ew,Baumann:2007zm}. When the power spectrum of scalar modes is enhanced, the production of these induced GWs becomes significant, which might be probed by GW detectors like interferometers or pulsar timing array. 
Therefore, the scalar-induced gravitational wave can provide crucial insights into our understanding of the early universe, especially during periods of inflation and reheating \cite{Espinosa:2017sgp,Kohri:2018awv}. 
%The main purpose of this paper is to study the impact of BIS on the generation of scalar-induced GWs.
Several mechanisms have been proposed in the literature for a possible amplification of primordial density perturbations and the generation of induced GWs at small scales, including near-inflection points \cite{Starobinsky:1992ts,Ivanov:1994pa,Di:2017ndc,Ballesteros:2017fsr,Mahbub:2019uhl,Ragavendra:2020sop,Pi:2022zxs,Domenech:2023dxx}, step-like changes \cite{Kefala:2020xsx,Inomata:2021tpx,Dalianis:2021iig,Cai:2021zsp},  small periodic structures \cite{Peng:2021zon,Cai:2019bmk}, rapid turns in multi-field inflation \cite{Fumagalli:2021mpc,Palma:2020ejf,Fumagalli:2020nvq}, sound speed resonances \cite{Cai:2018tuh,Cai:2020ovp,Addazi:2022ukh}; Higher-Dimensional Operators and Non-Minimal Coupling, including interactions between the inflaton and the graviton \cite{Kannike:2017bxn,Pi:2017gih,Capozziello:2017vdi,Capozziello:2019klx,Fu:2019ttf,Ashoorioon:2019xqc,Ashoorioon:2018uey,Lin:2020goi,Kawai:2021edk,Wang:2024vfv}; extra fields such as axion-like curvaton, spectator scalar fields or gauge bosons \cite{Biagetti:2013kwa,Cai:2021wzd,Ando:2017veq,Inomata:2020xad,Zhou:2020kkf,Pi:2021dft,Inomata:2022ydj,Abishev:2022wfe}; 
first-order phase transition \cite{Addazi:2018nzm,Kawana:2021tde,Baker:2021nyl,Liu:2021svg}; Supergravity and String-Inspired Models \cite{Aldabergenov:2020bpt,Aldabergenov:2020yok,Wu:2021zta,Zhang:2021rqs,Spanos:2021hpk,Ketov:2021fww}.

In this paper, we propose a new mechanism for enhancing the power spectrum of scalar perturbations through complex mass transient instabilities during inflation.
After inflation and reheating, 
the complex mass modes re-enter 
to the stability domain.
We will show that 
BIS transient instabilities
generate a characteristic class of blue-tilted GW spectra 
which can be tested 
by future space-based interferometers 
such as {\bf LISA}, {\bf DECIGO}, {\bf BBO}, {\bf Taiji}, {\bf TianQin} 
and terrestrial experiments 
like {\bf ET} and {\bf CE}. 
Implications for Primordial Black Holes (PBHs)
will be also explored and discussed. 

The paper is organized as follows. In the next section, we introduce the model under consideration and discuss the non-local extensions of the geometric Starobinsky inflation and its scalar field formulation in the Einstein frame. In Section III we formulate the quantisation and compute the scalar power spectrum of BIS. After that, we proceed with the computation of the corresponding scalar-induced GW signal. In Section IV we discuss the phenomenological implications of the presence of BIS-s during inflation for GWs and PBHs. In the last Section we summarise the results and discuss the perspective to probe and constrain quantum gravity with GW experiments and cosmological observations.

\section{The model}
\subsection{Geometric Starobinsky inflation}

Let us start with a nonlocal gravity action as follows \cite{Koshelev:2017tvv}: 
\begin{equation}
\label{S}
S=\int d^{4}x\sqrt{-g}\Big[\frac{M_{Pl}^{2}}{2}R+\frac{1}{2}R\mathcal{F}_{R}(\Box_s)R+\frac{1}{2}W_{\mu\nu\rho\sigma}\mathcal{F}_{W}(\Box_s)W^{\mu\nu\rho\sigma}\Big]\, , 
\end{equation}
where $W_{\mu\nu\rho\sigma}$ is the Weyl tensor and $\mathcal{F}_{R,W}\equiv \mathcal{F}(\Box_{s})$ are the form factors
with $(\Box_{s}/\mathcal{M}_{s})$ and $\mathcal{M}_{s}$ is the non-locality scale. 
We assume that $\mathcal{F}_{R,W}$ can be Taylor expanded at zero as 
\begin{equation}
\label{Taylor}
\mathcal{F}_{R,W}(\Box_{s})=\sum_{n=0}^{\infty}f_{nR,W}\Box_{s}^{n}\, ,
\end{equation}
\vspace{0.2cm}
with $f_{nR,W}$ the Taylor coefficients. 

A Lagrangian for a transverse and traceless graviton field around Minkowski space-time gets the form
\begin{equation*}
    L_2\sim \frac12 h_{\mu\nu}\Box(M_P^2-2\Box\mathcal{F}_W(\Box_s))h^{\mu\nu}\, . 
\end{equation*}
In general such a quadratic form results in various phenomena including complex squared mass poles. It is reasonable to worry that they might be ghosts.
Extra poles compared to a local theory can be avoided with proper constraints on the 
form factors. 
However, the ghost-free constraints are background-dependent: eliminating ghosts in Minkowski background does not 
guarantee that the theory is ghost-free in another background.
%We will come to this shortly.
Thus, in general, a ghost-free non-local theory in Minkowski space-time
can have extra complex mass squared scalars in de Sitter space-time.

Here is a brief explanation of the above statement. As it is known, the exponential of entire functions is the only class of functions that have no zeros on the whole complex plane 
and thus can be used to construct form factors that avoid ghosts. In general, they have a form
\begin{equation}
\label{exp}
\mathcal{F}_{R}(\Box_{s})=\frac{M_{Pl}^{2}}{6}\frac{1-(1-\Box/M^2)e^{\omega(\Box_{s})}}{\Box}
\end{equation}
and 
\begin{equation}
\label{exp2}
\mathcal{F}_{W}(\Box_{s})=\frac{M_{Pl}^{2}}{2}\frac{e^{\omega(\Box_{s})}-1}{\Box}\, .
\end{equation}
%Here we assume that a scalar degree of freedom is present. In an infinite derivative case we in principle can avoid an extra scalar but we want to have one to organize an inflation.
This leaves only a single perturbative degree of freedom in the spin-2 sector --- the massless graviton.
A factor $(1-\Box/M^2)$ results in the appearance of a scalar degree of freedom, i.e. scalaron, or inflaton, and in general can be absent in this setup.
%Note that it is not possible to have no scalar in just $f(R)$ gravity theory.
One and the same entire function $\omega(\Box_s)$ and factors $1/6$ and $1/2$ with exactly a ratio of $3$ are required to consistently modify a graviton propagator. Indeed, such form factors modify the graviton propagator as follows
\begin{equation}
\label{Pro}
\Pi(p^{2})\sim -\frac{P^{(2)}}{p^{2}e^{\omega(-p^{2})}}+\frac{P^{(0)}}{2p^{2}\Big(1+\frac{p^{2}}{M^{2}} \Big)e^{\omega(-p^{2})}}\, ,
\end{equation}
where $P^{(2),(0)}$ are spin projection operators. 

On the other hand, as mentioned above and shown explicitly in \cite{Biswas:2016egy} by computing a propagator in (anti-) de Sitter space
%, to avoid extra states the form-factors in any other background.  Apparently this is what we need to study inflation. 
%Sticking to a pure de Sitter
we can show that one can avoid ghosts fixing
\begin{equation}
\label{FC}
\mathcal{F}_{W}(\Box_{s})=f_{0}\bar{R}_{dS}\frac{e^{\gamma_{T}(\Box_{s}-\frac{\bar{R}_{dS}}{3\mathcal{M}_{s}^{2}})}-1}{\bar{\Box}_{dS}-\frac{2}{3}\frac{\bar{R}_{dS}}{\mathcal{M}_{s}}}\, , 
\end{equation}
where $\bar{R}_{dS}$ is the value of the Ricci 
scalar on a chosen de Sitter space. Explicit computations seem feasible only around a de Sitter space-time. One can see that, in order to avoid extra states in any background, the form-factors have to be background-dependent. This however may be in conflict with renormalizability claims (see \cite{Koshelev:2022olc}) as well as inflationary predictions. From this perspective, it looks more realistic to work with extra states around non-flat backgrounds.

Let us add several important considerations and remarks on these aspects. 
 First, if form factors have fixed constant coefficients, it will be not possible to have no extra states even for two beforehand chosen de Sitter spacetimes. Second, in the case of inflation, the de Sitter stage is approximate and such a tuning of form factors would be, anyway, useless. Third, it is worth to remark  that once the theory parameters are fixed, these cannot be re-adjusted depending on the any other specific situations. The latter implies that if 
 we choose conditions for no extra states in 
  Minkowski background, then we will encounter BIS around other backgrounds. 
  On the other hand, 
  keeping Minkowski space-time as in a local case seems to be the only reasonable assumption as complex masses squared are absent in a flat space-time.

Model (\ref{S}) allows for an exact embedding of the Starobinsky solution and scalar perturbations were studied in many details in \cite{Craps:2014wga} albeit without the Weyl tensor part.
The equation for a canonical variable was obtained to be an infinite derivative analog of a local counterpart having the form
\begin{equation*}
    \mathcal{G}(\Box_{s})(\Box_s-M^2)\chi=0
\end{equation*}
In \cite{Craps:2014wga} an exponential of an entire function form was assumed for $\mathcal{G}(\Box_{s})$ implicitly meaning extra states on the Minkowski background.
Here we specifically want to go on the other way preserving the Minkowski background and keeping only one massless graviton on it. This means that, considering the model (\ref{S}), 
we do not assume that $\mathcal{G}(\Box_{s})$ is an exponential of an entire function thereby infinitely many new BIS with complex squared masses.

%The second order action for scalar perturbations corresponds to 
%\begin{equation}
%\label{scalar}
%\delta^{(2)} S=\frac{1}{2f_{0}\bar{R}_{dS}}\int d^{4}x\sqrt{-g}\mathcal{Y}\frac{\mathcal{W}(\bar{\Box}_{s})}{\mathcal{F}_{R}(\bar{\Box}_{s})}(\bar{\Box}_{s}-M^{2}) \mathcal{Y}
%\end{equation}
%where $\mathcal{Y}$
%is the canonical variable 
%related to curvature pertrubation as 
%$\mathcal{Y}\sim f_{0}\bar{R}_{dS}\mathcal{R}$
%and 
%\begin{equation}
%\label{WW}
%\mathcal{W}(\Box_{s})=3\mathcal{F}_{R}(\Box_{s})+(\bar{R}_{dS}+3M^{2})\frac{\mathcal{F}_{R}(\Box_{s})-f_{0}}{\Box-%M^2}\, , 
%\end{equation}
%which, for Eq.\ref{exp} corresponds to
%\begin{equation}
%\label{WWW}
%\mathcal{W}(\Box_{s})=-f_{0}\bar{R}_{dS}\Big(\frac{1-e^{\gamma_{S}(\Box_{s})}}{\Box} \Big)+3f_{0}e^{\gamma_{S}(\Box_{s})}\, . 
%\end{equation}
%The zeros $z$ of $\mathcal{W}(z/\mathcal{M}_s)$
%correspond
%to solve the equation
%\begin{equation}
%\label{eee}
%\bar{R}_{dS}\Big( \frac{1-e^{\gamma_{s}(z/\mathcal{M}_{s}^{2})}}{z}\Big)=3e^{\gamma_{S}(z/\mathcal{M}_{s}^{2})}\, .
%\end{equation}
%In case $\gamma_{S}=\alpha_{1}\Box_{s}(\Box_{s}-M^{2}/\mathcal{M}_{S}^{2})$, one obtains  complex conjugate poles. 
%In case $M^{2}<<\mathcal{M}_{s}^{2}$, these poles approximately 
%correspond to $z_{\pm}=\pm \mathcal{M}_{s}^{2}\sqrt{n+1/2}(1\pm i)$
%where $n\geq 1$ is an integer. 

\subsection{Non-local scalar field inflation}

It is well-known that in the local case, geometric Starobinsky inflation is equivalent to the scalar field inflation with the specific choice of the potential. In non-local case, this equivalence is spoiled by an infinite number of additional couplings between the scalar field and curvature invariants in the Einstein frame action. But we can also start from the local action defined in the Einstein frame and embed it to UV-finite action for the scalar field, along the lines of Ref. \cite{Koshelev:2020fok}. In this work, we are dealing with non-local scalar field instead of a non-local geometrical action because, while we expect the same phenomena to take place in geometric formulation, the Einstein frame formulation is technically much simpler. In addition, 
it can be generalized to arbitrary inflaton potential 
in a simple way that is 
possible for geometric action (\ref{S}). 

Let us start with the following action,
\begin{equation}
S=\int d^4x\sqrt{-g}\left(\frac{M_P^2}2R+\frac12\varphi(\Box_{s}-M^2)f(\Box_{s})\varphi-v(\varphi)\right).
    \label{Smodel}
\end{equation}
To keep a standard IR limit, we should have a normalization $f(0)=1$ and a full potential if $V=\frac{M^2}2\varphi^2+v(\varphi)$ where $M$ is the inflaton mass. Here, we specifically separate the mass term from the 
$v(\varphi)$ and consider it as modified by 
 the form factor $f(\Box_{s})$. 
This choice guarantees
that $v(\varphi)$ does not contribute to the propagator and the absence of BIS in the Minkowski limit. A condition to have no extra state in Minkowski spacetime is 
\begin{equation}
f(\Box_{s})=e^{2\sigma(\Box_{s})}
\end{equation}
with $\sigma(0)=0$ for a proper normalization. Below we will set $\mathcal{M}_s=1$, i.e. $\Box=\Box_s$.

Let us consider fluctuations of the field $\varphi$ around a constant value $\varphi_{0}\neq0$ as
$\varphi=\varphi_{0}+\delta \varphi$.
The equation of motion for perturbations reads as 
\begin{equation}
\label{eeqq}
\Big(\Box f(\Box)+V''(\varphi_{0})\Big)\delta \varphi=0\, . 
\end{equation}
Unless $V''(\varphi_0)=0$ the latter operator cannot be an exponential of an entire function as long as $f(\Box)$ is.
However, in the exact de Sitter space, it can be factorized using the Weierstrass decomposition 
\begin{equation}
\label{crr}
\prod_{i=1}^{\infty}\Big(\Box-\mu_{i}^{2}\Big)\delta \varphi=0 \, .
\end{equation}
Each factor $(\Box-\mu_i^2)$ corresponds to a pole in the propagator for perturbations and thus can be associated with a particle. Given that $f(\Box)$ is an exponential any constant shift will result in infinitely many factors. This is a simple reflection of the fact that different backgrounds in gravity cannot be simultaneously free of extra states.
Indeed, we have
\begin{equation}
\label{Boxe}
\Big(\Box e^{\sigma(\Box)}+V''\Big)\delta\varphi=0,
\end{equation}
which corresponds to 
\begin{equation}
\label{loggg}
(\log \Box+2\sigma(\Box)+\log\, V''+2\pi i N)\delta\varphi_N=0\, .
\end{equation}
for any integer $N$, i.e. any $\delta\varphi_N$ is a solution to the original equation. This latter construction generates an infinite number of pairs of solutions corresponding to complex conjugate squared masses. 

The situation is much more complicated for a non-trivial background deviating from the maximally symmetric spacetime, as the Weierstrass product would not work. This question on its own is very difficult and has limited studies. During inflation, though, we can approximate $V''\approx\,$const which is expected to be small if slow-roll conditions are satisfied but it is definitely wrong to assume $V''=0$.
Complex conjugate poles like any other poles produce classical degrees of freedom.
A similar situation was also found for Lee-Wick theories \cite{Anselmi:2018kgz}. 
Such states are often simply ignored in any previously performed analysis by invoking new projection rules. 
But in principle, BIS-s can exist and they can cause different physical effects in cosmology. 

Let us make two comments about these aspects. The first is that, contrary to local scalar-tensor models, a slow-roll condition for inflation and Hubble parameter does not immediately imply any condition on the speed of the scalar field. This is because rigorously speaking, equations of motion contain a lot of terms coming from the variation of d'Alembertians inside the form-factor. One can however quite easily show that a standard slow-roll condition for a scalar field such that $\dot\varphi^2/(2M_P^2H^2)\ll 1$ is consistent as long $M\ll \mathcal{M}_s$. The second is an observation related to the first statement that, indeed a hierarchy $M\ll \mathcal{M}_s<M_P$ should be considered. Namely, inflation can happen without higher derivatives so that they should be only corrections to the background dynamics. On the other hand, these corrections are considered to be game-changers for gravity normalization in the quantum regime and thus $\mathcal{M}_s$ should not be higher than Planck mass, fixing the GR term scale.

Thus, we consider such new BIS with complex conjugate masses squared
within the assumptions aforementioned.
As the simplest example, we can start with activating just one pair of such states.
In the next Section, we will compute the power spectrum of scalar perturbations and scalar-induced GWs resulting from BIS. 
We will show how the complex conjugate pairs can contribute to the scalar power spectrum beyond standard local inflation.

\section{Power Spectrum of scalar perturbations}

The power spectrum of scalar perturbations can be obtained just considering the  scalar field sector of the model (\ref{Smodel}). 
This is a good approximation as long as the slow-roll conditions are satisfied: the contribution of metric perturbations to the curvature perturbation is suppressed by the slow-roll parameters, compared to the contribution of the inflaton field  \footnote{In addition, we have to demand that the energy scale of non-locality is much larger than the Hubble scale of inflation. Indeed, this condition has to be satisfied, in order to keep the background solution as the same as in the local case \cite{Koshelev:2020fok}.}. Therefore, we can study equation (\ref{eeqq}) which we re-write here for the sake of completeness as
\begin{equation}
\label{eeqqagain}
\Big(\Box f(\Box)+V''(\varphi_{0})\Big)\delta \varphi=0\, .
\end{equation}
As suggested at the end of the previous Section, let us consider one pair of BIS with $\mu_{\pm}^{2}=m_{1}^{2}\pm im_{2}^{2}$ 
on top of the inflationary background: 
\begin{equation}
\label{inflationa}
\delta \ddot{\phi}+3H\delta \dot{\phi}+\Big( \frac{k^{2}}{a^{2}}+\mu_{+}^{2}\Big)\delta \phi=0 \, , 
\end{equation}
and same for $\mu_{-}$.
We can substitute $\delta \phi$ in Eq.\ref{inflationa} with $\delta \phi=\chi/a$
obtaining 
\begin{equation}
\label{chii}
\chi''+\Big(k^{2}-\frac{a''}{a}+\mu^{2}_{+}a^{2} \Big)\chi=0
\end{equation}
with $a=-1/H\eta$ in terms of the conformal time.

Quantization of the fields with complex-conjugate masses in de-Sitter was performed in \cite{Tokareva:2024sct} (see also an earlier paper \cite{Yamamoto:1970gw,Yamamoto:1970di} where a flat space case was considered\footnote{We do not consider this case because in flat space complex mass fields cannot be classically stable which follows from \eqref{StabilityCondition} in the limit $H\rightarrow 0$.}) for the tensor perturbations. Here we repeat the same procedure for the case of the scalar field. Let us consider the Bunch-Davies vacuum condition 
\begin{equation}
\label{BD}
\eta \rightarrow -\infty,\,\,\,\, \chi^{+}=e^{ik\eta}\, . 
\end{equation}

The corresponding quantum field has two modes
\begin{equation}
\label{corr}
\hat{\chi}^{(i)}(\mu_{+})=\int \frac{d^{3}k}{\sqrt{2k_{0}}(2\pi)^{3/2}}\Big( \chi^{+}(\eta,\mu_{+})\hat{A}_{(i)}^{+}(k) e^{-i{\bf k}{\bf x}}+ \chi^{-}(\eta,\mu_{+})\hat{A}_{(i)}^{-}(k))e^{i{\bf k}{\bf x}}\Big)\, ,
\end{equation}
\begin{equation}
\label{corr2}
\hat{\chi}^{(i)}(\mu_{-})=\int \frac{d^{3}k}{\sqrt{2k_{0}}(2\pi)^{3/2}}\Big( \chi^{-}(\eta,\mu_{-})\hat{A}_{(i)}^{+}(k))e^{-i{\bf k}{\bf x}}+ \chi^{+}(\eta,\mu_{-})\hat{A}_{(i)}^{-}(k))e^{i{\bf k}{\bf x}}\Big)\, ,
\end{equation}
with 
\begin{equation}
\label{condd}
\chi^{-}(\eta,\mu_{-})=\Big(\chi^{+}(\eta,\mu_{+})|_{\mu_{+}\rightarrow \mu_{-}}\Big)^{*}\, . 
\end{equation}
Here the index $(i)$ labels the two physical states per each pair of the complex-conjugate masses. They can be also described by the quadratic action of two complex-valued fields with 2 constraint equations (required to make the action real-valued), as it is also presented in \cite{Koshelev:2020fok,Tokareva:2024sct}.

The total field including $\pm$ modes is 
\begin{equation}
\label{PPP}
\hat{\chi}=\int \frac{d^{3}k}{\sqrt{2k_{0}}(2\pi)^{3/2}}\Big(\chi^{+}(\eta,\mu_{+})+\chi^{-}(\eta,\mu_{-}) \Big)\hat{A}_{k}^{+}e^{-i{\bf k}{\bf x}}.
\end{equation}
The power spectrum of $\hat{\chi}$ reads as 
\begin{equation}
\label{con}
\mathcal{P}_{\chi}=\frac{k^{2}}{(2\pi)^{2}}\Big|\chi^{+}(\eta,\mu_{+})+\left(\chi^{+}(\eta,\mu_{+})|_{\mu_{+}\rightarrow \mu_{-}}\right)^{*}\Big|^{2} 
\end{equation}
and the corresponding power spectrum for $\hat{\phi}$ and scalar curvature perturbations are
\begin{equation}
\label{cooo}
\mathcal{P}_{\phi}=\frac{H^{2}k^{2}\eta^{2}}{(2\pi)^{2}}\Big|\chi^{+}+(\chi^{+}|_{\mu_{+}\rightarrow \mu_{-}})^{*}\Big|,\,\,\,\ \mathcal{P}_{R}=\frac{H^{2}}{\dot{\phi}^{2}}\mathcal{P}_{\phi}\, . 
\end{equation}
Strictly speaking, these fields correspond to the isocurvature perturbations, rather than curvature perturbations. However, under the reasonable assumption that reheating happened instantaneously right after inflation, and these fields decayed to the SM radiation with the same rate as the inflaton field, these isocurvature perturbations are transformed to curvature ones. We will return to this point later.

The solution of EoM with BD initial conditions is 
\begin{equation}
\label{chipm}
\chi^{+}=\frac{1}{2}(i+1)e^{\frac{\pi}{4}i\sqrt{9-4\nu_{+}}}\sqrt{-\pi k \eta}\Big[\mathcal{J}_{\frac{1}{2}\sqrt{9-4\nu_{+}}}(k\eta)+i\mathcal{Y}_{\frac{1}{2}\sqrt{9-4\nu_{+}}}(k\eta)\Big]\, ,
\end{equation}
where $\nu_{+}=\mu_{+}^{2}/H^{2}$. 

Let us remark that the bound condition for no growing modes during inflation corresponds to $m_{2}^{4}<9 m_{1}^{2}H^{2}$. It dynamically changes during inflation since $H$ is not constant. Therefore, some stable modes can exit the stability condition becoming temporarily unstable.

Eqs. \eqref{chipm} and \eqref{cooo} can be further simplified, and lead to the following expression for the power spectrum,
\begin{equation}
\label{PRR}
P_{R}=\frac{H^4 k^2\eta^2}{(2\pi)^2\dot{\phi}^2}\,(-\pi k\eta)\,\left|e^{\frac{i\pi}{4}\sqrt{9-4\nu_+}}\mathcal{H}^{(1)}_{\frac{1}{2}\sqrt{9-4\nu_{+}}}\,(-k\eta+ i\varepsilon)+e^{-\frac{i\pi}{4}\sqrt{9-4\nu_+}}\mathcal{H}^{(2)}_{\frac{1}{2}\sqrt{9-4\nu_{+}}}\,(-k\eta-i \varepsilon)\right|^2\, 
\end{equation}
Apparently, we note that Eq.\ref{PRR} provides better accuracy for numerical algorithms implemented in {\bf Wolfram Mathematica}, especially for large imaginary orders of the Hankel function. Here $\varepsilon>0$, $\varepsilon\rightarrow 0$ stands for selecting the specific branch of Hankel functions. Let us note that all quantities in this expression (including $\nu_+$) should be taken at the moment of horizon exit for each mode. In this way, the power spectrum is slightly dependent on the inflaton potential. 

Let us define $\eta_{e}$ as the conformal time corresponding to the end of inflation: the $\epsilon(\eta_{e})\simeq 1$ where $\epsilon\equiv \frac{M_{P}^{2}}{2}\Big( \frac{V_{\phi}}{V}\Big)^{2}$. For the mode $k$, the moment of horizon exit corresponds to $k\eta_{k}=1$ and $N_{k}=\log \eta_{k}/\eta_{e}$ is the e-fold number, and $N_{k}=M_{P}^{-2}\int_{\phi_{k}}^{\phi_{e}}\frac{V}{V_{\phi}}d\phi$, which can be inverted as $\phi_{k}(N_{k})$. The Hubble rate for the k-mode can be found as $H_{k}^{2}=(3M_{P}^2)^{-1}V(\phi_{k})$ at $\eta=\eta_{k}$. For a given potential one can find $\phi_{k}(N_{k})$ and $H_{k}(N_k)$ as we will show in the next section for Starobinsky's potential.

%In the limit of $\eta\rightarrow 0$, the leading order of the power spectrum corresponds to 
%\begin{equation}
%\label{jjal}
%|\chi^{+}+\chi^{-}|^{2}\Big|_{\eta\rightarrow 0}=C(k,H)(k\eta)^{\alpha(H)}\cos\Big(\beta \log k\eta+\phi_{0}(k)\Big)\, 
%\end{equation}
%which corresponds to 

\subsection{The case of Starobinsky inflation and generalizations}

Let us now consider Starobinsky's inflation:
\begin{equation}
\label{spec}
V=V_{0}\Big(1-e^{-\phi/\Lambda} \Big)^2\, , 
\end{equation}
where $\Lambda=\sqrt{6}M_{P}/2$ in case of $R+\gamma R^2$-gravity, and $V_0=3 H_0^2 M_P^2$, where $H_0=6.5\cdot10^{-6}\, M_P$ is demanded for CMB normalization.

In this case, the $\epsilon$ parameter corresponds to 
\begin{equation}
\label{epi}
\epsilon=\frac{2M_{P}^{2}}{\Lambda^2}\frac{1}{e^{\phi/\Lambda}-1}\, . 
\end{equation}
Thus, $\epsilon=\epsilon_e \sim 1$ corresponds to $\phi_{e}\sim \Lambda \log \Big(1+\Lambda^{2}/2M_{P}^2\, \epsilon_e \Big)$. 
Therefore, 
\begin{equation}
\label{Nks}
N_{k}=\frac{\Lambda}{2M_{P}^2}\int_{\phi_{e}}^{\phi_{k}} \Big(e^{\phi/\Lambda}-1\Big)d\phi=\frac{\Lambda^{2}}{2M_{P}^{2}}e^{\phi/\Lambda}\Big|_{\phi_{e}}^{\phi_{k}}+\frac{\Lambda}{M_{Pl}^2}(\phi_{e}-\phi_{k})\, \simeq \frac{\Lambda^{2}}{M_{P}^2}e^{\phi_{k}/\Lambda}\, \,\,\, (e^{\phi_{k}/\Lambda}\gg e^{\phi_{e}/\Lambda})\, . 
\end{equation}
Inverting this equation, we obtain 
\begin{equation}
\label{phikkk}
\phi_{k}=\Lambda \log \frac{2M_{P}^{2}N_{k}}{\Lambda^{2}}\, , 
\end{equation}
and, inserting it into the generalized Starobinsky potential \eqref{spec}, in the Hubble rate and in the derivative in time of the field, we obtain
\begin{equation}
\label{obt}
V(\phi_{k})=V_{0}\Big(1-\frac{\Lambda^2}{2M_{P}^2 N_k} \Big)^{2},\,\,\,\, H_{k}=\frac{V(\phi_{k})}{3M_{P}^2}=H_{0}\Big(1-\frac{\Lambda^2}{2M_{P}^2\, N_k} \Big)\,,\,\,\, \dot{\phi}_{k}=-\frac{3V_{\phi}(\phi_{k})}{3H_{k}}=\frac{\Lambda H_{0}}{N_{k}}\, . 
\end{equation}
The power spectrum corresponds to
\begin{equation}
\label{power1}
\mathcal{P}_{R}=\frac{H_{k}^4}{\dot{\phi}_{k}^{2}(2\pi)^2}(k\eta_{e})^2\Big(|\chi^{+}+\chi^{-}|^{2}\Big|_{H=H_{k},\eta=\eta_{e}}\Big)\, ,
\end{equation}
or
\begin{equation}
\label{PRNl}
\mathcal{P}_{R}(N_{k})=\frac{H_{0}^{2}\Big(1-\frac{\Lambda^{2}}{2M_{P}^{2}N_{k}} \Big)^{4}N_{k}^{2}}{(2\pi)^{2}\Lambda^{2}}e^{2N_{k}}|\chi_{+}+\chi_{-}|^{2}\, . 
\end{equation}
Thus, for the standard inflaton perturbations, we obtain the power spectrum as
\begin{equation}
\label{PR0}
\mathcal{P}^{(0)}_{R}(N_{k})=\frac{H_{0}^{2}\Big(1-\frac{\Lambda^{2}}{2M_{P}^{2}N_{k}} \Big)^{4}N_{k}^{2}}{(2\pi)^{2}\Lambda^{2}}\, . 
\end{equation}

The final expression for the power spectrum of the BIS is
\begin{equation}
\label{PRfinal}
\begin{split}
   {\cal P}_{R}= \frac{H_0^2 e^{-3 N_k} }{128 \pi  \Lambda ^2 N_k^2
   M_P^8} (\Lambda ^2-&2 N_k M_P^2 )^4 \left| e^{\frac{1}{4} i \pi  \sqrt{9-\nu _+(N_k)}}\mathcal{H}_{\frac{1}{2}\sqrt{9-\nu
   _+(N_k)}}^{(1)}\left(-e^{-N_k}+i \varepsilon\right)+\right.\\
   &\left.+e^{-\frac{1}{4} i \pi  \sqrt{9-\nu _+(N_k)}} 
   \mathcal{H}_{\frac{1}{2}\sqrt{9-\nu
   _+(N_k)}}^{(2)}\left(-e^{-N_k}-i \varepsilon\right)\right|^{\,2}\, ,
   \end{split}
\end{equation}
where $\nu_{+}(N_k)$ should be taken at the moment of horizon exit for the mode $k$, hence
\begin{equation}
    \nu_{+}(N_k)=\frac{16 \left(m_1^2+i \,m_2^2\right) N_k^2 M_P^4}{H_0^2 \left(\Lambda ^2-2 N_k M_P^2\right){}^2}\,.
\end{equation}

For large values of $m_1/H_0$ we can approximate the order of Hankel functions as
\begin{equation}
    \frac{1}{2}\sqrt{9-\nu_+}\approx-\frac{m_2^2}{2 H_0 m_1}+i \frac{m_1}{H_0}
\end{equation}
The expression in the modulus brackets in\eqref{PRfinal} scales as 
\begin{equation}
   \left| e^{\frac{1}{4} i \pi  \sqrt{9-\nu _+(N_k)}}\mathcal{H}_{\frac{1}{2}\sqrt{9-\nu
   _+(N_k)}}^{(1)}\left(-e^{-N_k}+i \varepsilon\right)+\right.
   \left.+e^{-\frac{1}{4} i \pi  \sqrt{9-\nu _+(N_k)}} 
   \mathcal{H}_{\frac{1}{2}\sqrt{9-\nu
   _+(N_k)}}^{(2)}\left(-e^{-N_k}-i \varepsilon\right)\right|^{\,2}\, \propto \frac{m_1^2}{H_0^2} e^{\alpha N_k},
\end{equation}
where $\alpha=-m_2^2/(m_1 H_k)$. This asymptotic can be obtained from the small argument and large (imaginary) index expansion of Hankel functions. We have also verified this scaling numerically. Recall that $N_k$ can be related to $k$ as $N_k=-\log{k/k_{end}}=N_e-\log{k/k_*}$ where $k_{end}$ corresponds to the mode exiting horizon at the end of inflation, $k_*$ is a Planck pivot scale. The growing part of the power spectrum can be approximated as a power-law spectrum with the following parametric dependence,
\begin{equation}
\label{tilt}
    {\cal P}_R(k)\propto  \left(\frac{k}{k_*}\right)^{\omega}, \quad \omega=3+\alpha - \frac{2}{N_k}=3-\frac{m_2^2}{m_1 H_k} - \frac{2}{N_k}.
\end{equation}
Here the contribution $-2/N_k$ is the well-known spectrum tilt for the curvature perturbation in the Starobinsky model. The spectral index is slightly growing with $k$ but as the first approximation, one can take $H_k$ at $N_k\approx 55$ in \eqref{tilt} corresponding to the CMB mode. We found that it provides a good estimate for the upper bound for the spectral tilt, while the lower bound can be obtained if $H_0$ is taken instead of $H_k$. For more accurate results for the whole spectrum the full expression \eqref{PRfinal} should be used. In what follows we do numerical computations for the GW spectrum with the use of \eqref{PRfinal} for the scalar power spectrum. We found that for the benchmark parameters presented in Table 1, the spectrum tilts $\omega$ are in the range between $\omega=0.35$ and $\omega=0.65$.
%\sout{The steeper growth is possible but it should start at higher momenta in order to avoid a strong coupling regime.  However, it brings the predictions for the GW spectrum to the area that cannot be probed in future observations. The smaller spectral index is possible but it also doesn't lead to any observable effects.} 
Steeper growth is possible, which should start at higher momenta in order to avoid strong coupling. However, it brings predictions for the GW spectrum to undetectable regimes even in the future. Similarly, milder growth is also possible, but it does not lead to observable effects.

The total scalar power spectrum will get contributions from all complex-mass BIS. Strictly speaking, their impact on the curvature mode (which is related to the density contrast in the late Universe and is measured by {\bf Planck}) might depend on the details of the preheating stage. However, here we are reasonably assuming that these states (being parts of the same inflaton field) decay synchronously with the same rates as the SM particles leading to instantaneous reheating. Under this assumption, we can compute the total power spectrum as a direct sum of the power spectra of each mode, including the standard (massless) inflaton perturbation. This allows us to get the final scalar power spectrum of non-local inflation. 

We find that the power spectrum gets an enhancement for modes with momenta higher than CMB scale, and this contribution is dominated by the modes that are close to the boundary of the stability region (see Figure 1). The stable states with complex masses that are far away from the parabola of stability give negligible effects on the total power spectrum, similar to the ordinary heavy fields. Thus, if just one of the infinite number of states appears near the parabola, it can lead to potentially observable effects caused by the enhancement of higher momenta.

%with 
%\begin{equation}
%\label{powa2}
%|\chi^{+}+\chi^{-}|^2=(-k\eta)^{1-\alpha}\Gamma\Big(\frac{1}{2}(\alpha-i\beta)\Big)\Gamma\Big(\frac{1}{2}(\alpha+i\beta)\Big)\frac{1}{\pi}2^{\alpha}\Big(\cosh \frac{\beta\pi}{2}+\sinh\frac{\alpha \pi}{2} \Big)\, . 
%\end{equation}
%where 
%\begin{equation}
%\label{roooo}
%\alpha=\frac{1}{\sqrt{2}}\sqrt{ 9- \frac{4m_{1}^{2}}{H^2}+\sqrt{\Big( 9- \frac{4m_{1}^{2}}{H^2}\Big)^2+\frac{16m_{2}^4}{H^4}  } },\,\,\, \beta= \frac{1}{\sqrt{2}}\sqrt{ -9+\frac{4m_{1}^{2}}{H^2}+\sqrt{\Big( 9- \frac{4m_{1}^{2}}{H^2}\Big)^2+\frac{16m_{2}^4}{H^4}  } }\,. 
%\end{equation}

\subsection{Scalar-induced gravitational waves}

Scalar perturbations can source second-order tensor perturbations through the nonlinear interactions \cite{Matarrese:1997ay}. The production of induced GWs is important when the scalar power spectrum is enhanced. 
Such dynamics can be important in the early universe as the energy density of scalar perturbations sources the GW background \cite{Ananda:2006af}. See also \cite{Domenech:2021ztg} for a review of these aspects. 

The energy density of scalar-induced GWs, \(\Omega_{\text{GW}}\), can be related to the enhanced scalar power spectrum through a complex transfer function that encapsulates the dynamics of the universe's expansion and the detailed physics of the perturbations \cite{Kohri:2018awv}. The GW energy density spectrum can be expressed as:
\begin{equation}
\label{OmegaGW}
\Omega_{\text{GW}}(k) =\frac{c_g}{6} \Omega_{rad}   \int_0^\infty dv \int_{|1-v|}^{1+v} du \left( \frac{4v^2 - (1+v^2-u^2)^2}{4vu} \right)^2 {\cal P}_{\mathcal{R}}(uk) {\cal P}_{\mathcal{R}}(vk)\, , 
\end{equation}
where \(P_{\mathcal{R}}(k)\) is the scalar power spectrum, and \(u\) and \(v\) are integration variables representing the interaction of different wave modes \cite{Baumann:2007zm,Mangilli:2008bw}. We use $\Omega_{rad}=2.47\cdot 10^{-5}$ for the current radiation energy fraction, and $c_g=0.4$ in the particle physics below the scale of inflation is described by the Standard model. This expression was obtained under the assumption of immediate reheating taking place right after inflation. However, in general, the relationship between the power spectrum enhancement and the GW signal is influenced by the reheating phase following inflation.  The specifics of this process, including its duration and dynamics, impact on the resulting GW spectrum \cite{Bartolo:2018evs}.

In the case of instantaneous reheating, where the transition is rapid and the universe quickly enters the radiation-dominated phase, a relatively straightforward expression can describe the GW energy density \cite{Kohri:2018awv}. Assuming the power-law shape of the scalar power spectrum, the GW spectrum simplifies, as the transfer functions involved are less complex, to a more direct correlation between the enhanced scalar power spectrum and the GW signal as
\begin{equation}
\label{OmegaGW2}
\Omega_{\text{GW}} h^2 \approx A_{\text{GW}} \left( \frac{k}{k_*} \right)^{n_{\text{GW}}} {\cal P}_{\mathcal{R}}^2(k_*)\, .
\end{equation}
Here \(A_{\text{GW}}\) is a normalization constant, \(k_*\) is a pivot scale, and \(n_{\text{GW}}\) represents the spectral index of the GW background \cite{Inomata:2019zqy,Inomata:2019ivs}. This expression highlights that the induced GW energy density is proportional to the square of the scalar power spectrum, demonstrating how enhancements at large \(k\) significantly boost the GW signal.

For power-law dependence of the power spectrum ${\cal P}_{\mathcal{R}}(k)$ the spectrum index is related to the one of the curvature spectrum $\omega$ given in \eqref{tilt} in a very simple way \cite{Inomata:2019zqy,Inomata:2019ivs,Pi:2020otn}
\begin{equation}
    n_{GW}=2 \omega\, ,
\end{equation}
if $n_{GW}$ is not larger than $\sim3$ \cite{Cai:2019cdl}. 
For potentially observable signals we expect $0.7\lesssim n_{GW}\lesssim 1.2$. GW spectra while $n_{GW}$ out of this range are difficult to be probed even in the future GW detectors.

%\subsection{Observational Implications}

%Detecting these scalar-induced GWs offers a window into the primordial universe's conditions and the nature of inflation. Current and future GW detectors, such as LIGO, Virgo, and space-based observatories like LISA, aim to probe these signals \cite{Abbott2016}. A positive detection would not only validate aspects of inflationary theory but also provide constraints on the power spectrum at small scales, shedding light on the physics of the early universe \cite{Caprini2016}.

%In summary, the enhancement of the power spectrum at large \(k\) leads to a significant production of scalar-induced gravitational waves, with their properties intricately linked to the details of reheating. Instantaneous reheating simplifies this relationship, allowing for clearer predictions of the GW signal based on the enhanced scalar power spectrum \cite{Kohri2018}.

\section{Results and discussion}

\subsection{Scalar power spectrum and secondary gravitational waves.}

We present a computation of the power spectrum and the corresponding secondary GW signals for several benchmark points (see Table 1 and Figure 1) in the space of complex-valued masses. We choose such values corresponding to a real part of the mass around the Planck scale and the imaginary part close to the edge of the stability region. Our plots show that for these choices of parameters, the power spectra and the GW signals lie in the region accessible to future observational probes.

In particular, we found that the spectrum has a power-law shape rising for larger values of momenta (or for higher frequencies of GWs) up to a UV cutoff with a position and form that is strongly dependent from the details of preheating. Let us remark that this uncertainty does not affect the spectrum at lower momenta. This means that our predictions for the GW frequencies testable for space-based detectors are robust. However, in order to avoid too large values of the power spectrum at the end of inflation, we have to assume that BIS decay early enough. We found that if the effect of these modes starts to decrease about 10 e-foldings before the end of inflation (see the last line in Table 1) they do not produce too many PBHs which could evaporate leading to contradiction with the observations. However, given that it is possible that non-perturbative gravity would not support singular "standard" solutions, these constraints could be relaxed. As the dynamics of PBH formation in non-local gravity is an extremely complicated and non-fully explored subject beyond the scope of the current paper, we can also assume here that the final power spectrum does not lead to PBH production. Thus, we will mainly refer to the solid lines in the plots for the power spectrum and GWs, as they are phenomenologically reasonable and of interest for future tests.

%\textcolor{red}{[Comment for us, AA: since we do not understand PBH formation and evaporation in non-local gravity $100\%$, it is lecite to consider scenarios where standard bounds are relaxed. Anyway it may be interesting to think about together for next projects... }

\begin{table}[]
    \centering
    \setcellgapes{5pt}\makegapedcells
    \begin{tabular}{|c||c|c|c|c|} \hline 
           & B1 & B2 & B3 & B4\\ \hline 
            \(\displaystyle ~\frac{{\rm Re~} m^2}{M_P^2}\) & 0.42 & 0.42 & 
     0.085 & 0.42 \\ \hline 
  \(\displaystyle~\frac{{\rm Im~} m^2}{M_P^2}\) & $~~1.06\cdot 10^{-5}~~$ & $~~1.01\cdot 10^{-5}~~$ & $~~4.8\cdot 10^{-6}~~$ & $~~9.7\cdot 10^{-6}~~$\\ \hline
 \(\displaystyle~\frac{{\rm Im~}m^2}{3 H_0\sqrt{{\rm Re~}m^2}}\) & 0.833 & 0.801 & 0.847 & 0.766 \\\hline
 $~$ Cutoff in $N_e~~$ & 11 & 9 & 9 & 6 \\\hline
 \end{tabular}
    \caption{Benchmark points (B1-B4) for computation of the power spectrum and GW signal. The last row corresponds to the moment around the end of inflation at which we assume that the growth of the mode is finished. We assume it happens because back-reaction effects and reheating dynamics came into the game. The numbers are given in terms of e-foldings left until the end of inflation.}
    \label{tab:my_label}
\end{table}
  \begin{figure}[h!]
        \centering
        \includegraphics[width=4in]{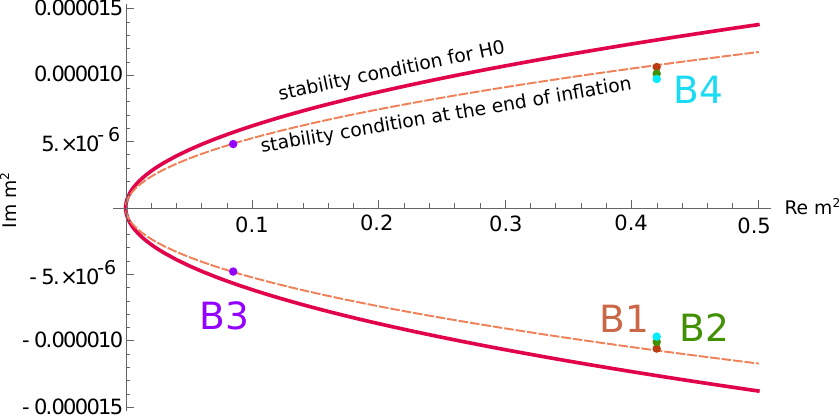}
        \caption{The positions of the benchmark parameters (TABLE I), concerning the stability condition for complex mass states, are displayed.}
        \label{parabola}
    \end{figure}

The power spectrum starts to rise at a certain critical value of momentum which strongly depends by the distance from the benchmark point to the stability parabola (see Figure 1 and Figure 2). A very tiny displacement from the boundary of the stability region triggers 
the start of the spectral growth at much higher frequencies.
Indeed, the points B1, B2, B4 are very close to each other but they lead to very different results, as shown in Figures. 

For lower momenta, the spectrum is exactly the same as in the local Starobinsky model. Thus our results fit the {\bf Planck} constraints   while having the growth of the spectrum at higher frequencies. The latter can be also probed in large-scale structure measurements sensitive to power spectrum shape at higher momenta.

%https://arxiv.org/pdf/2304.02996

    \begin{figure}[h!]
        \centering
        \includegraphics[width=6in]{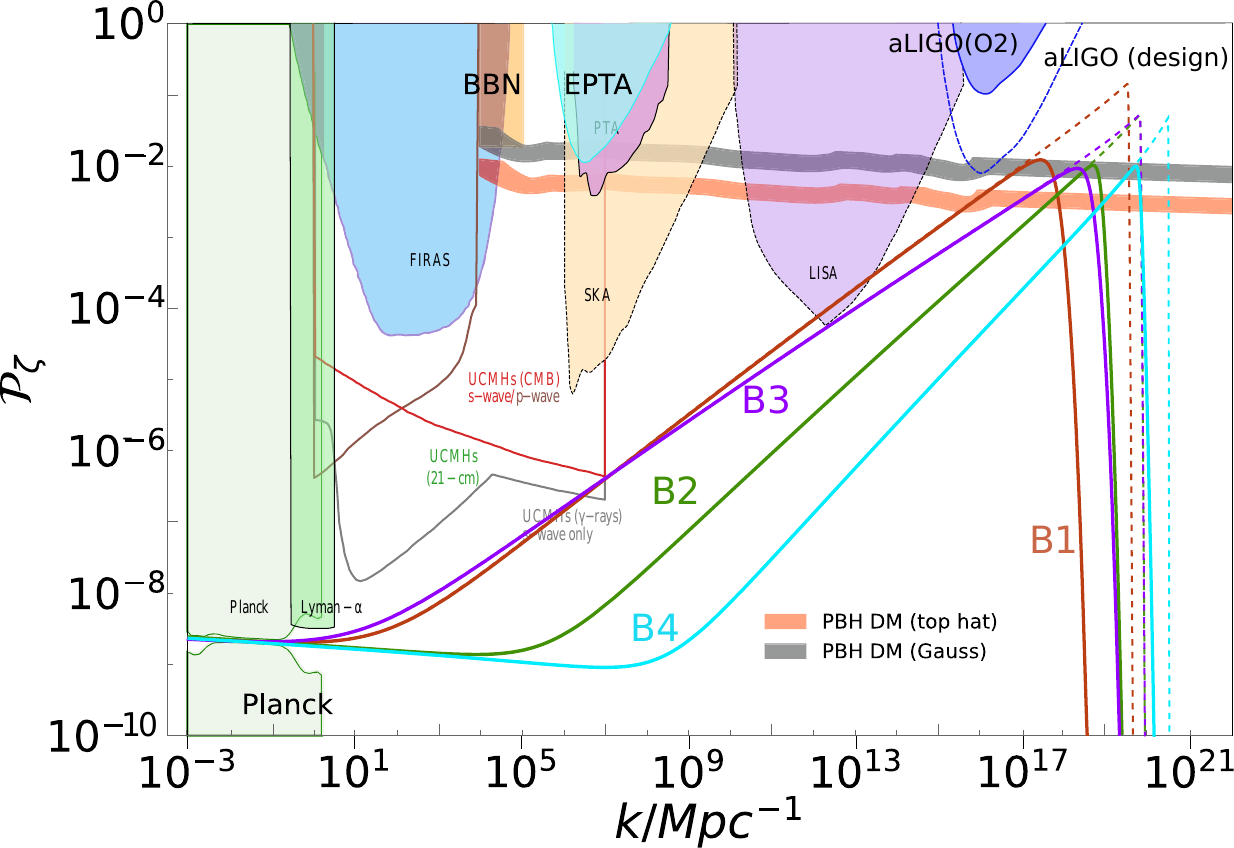}
        \caption{Plots of the power spectra generated by the complex mass modes. Solid lines correspond to the safe choice of the cutoff of the growth of the modes which does not lead to the overproduction of PBH (or PBH-like objects). The dashed line represents the choice of this cutoff moment closer to the end of inflation. This choice is in tension with the overproduction of PBH if this theory allows for their existence.
        Our results are compared with constraints (see Ref.\cite{FrancoAbellan:2023sby}) obtained independently from Planck \cite{Planck:2018vyg} (light green), Lyman-$\alpha$ data \cite{Bird:2010mp}  (darker green), FIRAS \cite{Fixsen:1996nj,Chluba:2012we}(blue), PTA  (pink); SKA
forecasts (yellow); LISA forecasts (purple) \cite{Inomata:2018epa}; UCMH constrains from free-free emission scenario \cite{Abe:2021mcv}, 21-cm \cite{Furugori:2020jqn} and diffuse $\gamma$-rays \cite{Delos:2018ueo}.  }
        \label{plotpr}
    \end{figure}

    \begin{figure}[h!]
        \centering
        \includegraphics[width=5in]{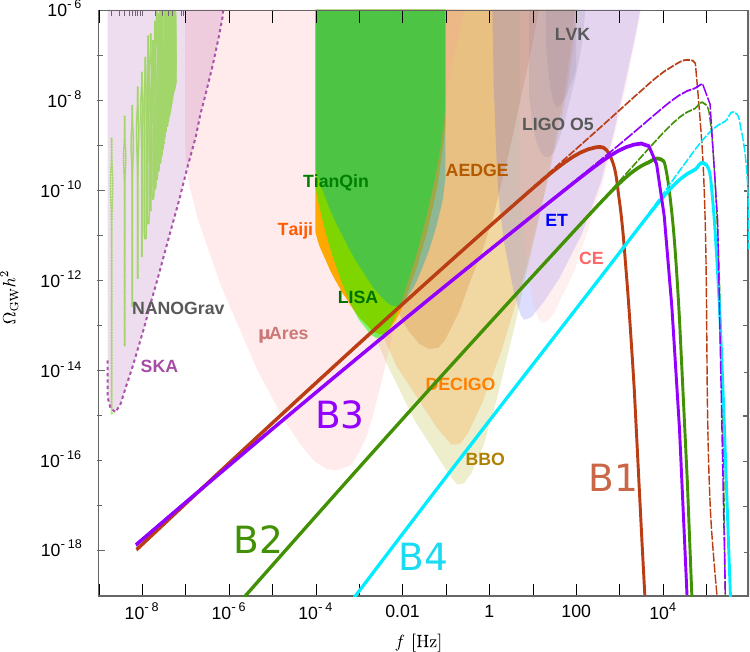}
        \caption{GW signal predictions for the 4 choices of the parameters (Table I and Figure I) are displayed in comparison with 
        the sensitivity curves of several future experiments, within the $(\Omega_{GW}h^{2},f)$ space. Solid lines correspond to the reasonable assumption that the growth of the modes is stopped by the back-reaction effects and the beginning of reheating, such that the power spectrum does not grow to a non-linear regime around the end of inflation. The model-dependent uncertainty about the mode behavior around the end of inflation (solid lines vs dashed lines) does not affect the predictions for GW spectrum at lower frequencies available to several future GW detectors. The corresponding sensitivity curves were taken from \cite{Marfatia:2023fvh}.}
        \label{plotgw}
    \end{figure}

What happens for the masses well below the Planck scale (for real parts of the mass)? We noticed that in this case, observable GW signals are possible only if the power spectrum rises to large values (larger than 1) at high momenta. For this reason, the allowed parameters of the theory correspond to unobservable features in the power spectrum. Interestingly, the observable signals refer to the states with real parts of the masses around the Planck scale.

Figure 3 shows our results for GW signals from temporary instability related to the presence of the complex mass states. 
The signals, in the frequency range tested by future space-based GW detectors, have a simple power-law form and their amplitude can be reached by {\bf LISA} \cite{LIGOScientific:2016wof} and {\bf Taiji} \cite{Ruan:2018tsw} experiments.  It is interesting to stress here that this type of power-law signal can be directly related to the effects of quantum gravity at the Planck scale.

\subsection{More about PBHs}

As it is known, the peak of $\mathcal{P}_{\zeta}(k)$ determines the PBH mass $M_{PBH}$ and ratio of the total amount of Cold Dark Matter (CDM) $f_{PBH}=\Omega_{PBH}/\Omega_{CDM}$. It is also well-known that the PBH mass is directly proportional to the cosmological time scale where their formation \cite{Carr:2020xqk} occurs as 
\begin{equation}
\label{PBH1}
M_{PBH}\sim \frac{c^{3}t}{G_{N}}\sim 10^{5}\Big(\frac{t}{10^{-23}s}\Big)\, g\, .
\end{equation}

%$$\beta( M )
%		&\equiv
%					\frac{ M\,n_{\rm PBH}( t_{i} ) }{ \rho( t_{i} ) }
%		\approx
%					7.98 \times 10^{-29}\,
%					\gamma^{-1/2}
%					\left(
%						\frac{ g_{* i} }{ 106.75 }
%					\right)^{\!1/4}
%					\left(
%						\frac{ M }{ M_{\odot} }
%					\right)^{\!3/2}
%					\left(
%						\frac{ n_{\rm PBH}( t_{0} ) }{ 1\,{\rm Gpc}^{-3} }
%					\right)
%					,
%					\label{eq:betaf}
%$$
%{\it where $\rho( t_{i} )$ is the density at time $t_{i}$ and $\gamma$ is the ratio of the PBH mass to the horizon mass. $g_{* i}$ is the number of relativistic degrees of freedom at PBH formation, normalised to its value at $10^{-5}\,\srm$ since it does not increase much before that in the Standard Model and this is the period in which most PBHs are likely to form. }

Assuming a standard PBH evaporation time-scale, this leads to a bound on $(f_{PBH},M_{PBH})$ from cosmic rays; more specifically from the extra-galactic $\gamma$-ray background \cite{Carr:2016hva}, the positron flux from Voyager \cite{Boudaud:2018hqb}, radiation from the Galactic bulge \cite{Laha:2019ssq}.

Thus, typically GW Power spectra corresponding to $k$-peak laying on the sensitivity curves of terrestrial interferometers such as {\bf LIGO} may correspond to too small mass and too large abundance of PBHs, excluded by evaporation bounds. 
Hawking's evaporation time $\tau\sim M_{PBH}^{3}$ will be smaller than the age of Universe if $M_{PBH}< 0.5\times 10^{15}\, {\rm g}\simeq 10^{-17}\, M_{\odot}$ or so. 

One of the open windows for PBH
corresponds to $M_{PBH}\sim 10^{-12}M_{\odot}$
with GW frequency peak around $mHz$ (LISA-scale).
In our case, the GW peak is always higher than 
$100\, {\rm Hz}$, related  
to $M_{PBH} \leq 10^{-17}M_{\odot}$ or so.

The following potential solutions are proposed:
i) PBHs or Mimickers, which are non-singular, horizonless exotic compact objects (ECOs), may form significantly later than predicted by the standard collapse model, possibly due to non-perturbative effects from non-local quantum gravity, evading ECO instabilities \cite{Addazi:2019bjz}. This could result in the formation of the last trapped surfaces occurring much later.
ii) The "standard" Hawking evaporation rate could be suppressed by new degrees of freedom, leading to an anomalous growth of the horizon, as discussed in the context of f(R)-gravity \cite{Addazi:2016prb}. This suppression could be associated with violations of the no-black-hole remnants conjecture \cite{Aharonov:1987tp,Banks:1992is,Susskind:1995da} or the Weak Gravity \cite{Arkani-Hamed:2006emk} and MSS conjectures \cite{Addazi:2021pty,Addazi:2023pfx}. Additionally, if black holes were replaced by a class of non-evaporating ECOs, such as Boson Stars \cite{Visinelli:2021uve}, then evaporation constraints would need to be reassessed.

\section{Conclusions}

\label{sec:Discussion-and-Conclusion}

In this work, we studied the effects of UV non-locality in gravity and inflaton sectors on the scalar power spectrum and the corresponding scalar-induced GW spectrum. We elaborate on the fact that the non-local propagator of the inflaton field inevitably results in the appearance of an infinite number of complex mass states when a non-trivial background is considered. This occurs if the condition of having no extra degrees of freedom in the true vacuum corresponding to flat space is satisfied. 

Naively, the BIS would lead to instabilities. However, around nearly de-Sitter space we can formulate a condition on the complex masses that guarantees the absence of infinitely growing perturbations in the superhorizon regime. This bound weakens as the Hubble scale of inflationary de Sitter decreases. As a result, some complex mass states may undergo a temporary instability that persists until the end of inflation. We show that this effect can lead to a rich phenomenology, including the formation of PBHs and secondary GW signals, due to the growing scalar power spectrum at large momenta.
Remarkably, we found that the potentially observable effects are associated with the real parts of the masses near the Planck scale
 which may naturally arise in UV completions of gravity within field theory, particularly in the non-local Starobinsky inflation model.

From a phenomenological perspective, 
we obtain GW spectra which exhibit distinct
features compared to other mechanisms previously analyzed in the literature. 
For instance, GW signals obtained cross
several orders of magnitude in frequencies 
with cross-correlated implications for {\bf $\mu$Ares}
in the $10^{-2}\div 1$ mHz window; for 
future space-based experiments 
such as {\bf LISA}, {\bf TianQin}, {\bf TAIJI} around the 
mHz; {\bf AEDGE}, {\bf DECIGO} and {\bf BBO} from mHz to Hz;
 for future  terrestrial interferometers such as {\bf ET} and {\bf CE} in the 1-100 Hz window.
Some of the solutions that we found 
do not violate any bounds on PBH overproduction; on the other hand, 
we also discussed a possible way out
from PBH Hawking's evaporation limits 
inspired by non-local quantum gravity. 
On the other hand, no GW spectra 
testable in Pulsar Timing array
were obtained without contradicting bounds from 
{\bf Planck} and Lyman-$\alpha$ data. 
In addition, we
also showed possible implications in Ultra Compact Mini-halos (UCMH)
free emissions, 21-cm and diffuse $\gamma$-rays 
physics as genuine tests of new physics through multi-messenger observations of non-local quantum gravity. 
Therefore, our analysis paves the way for several promising new methods to test infinite derivative quantum gravity and complex mass fields, even with mass parameters nearing the Planck energy scale.
%\sout{As a takeaway message of our work, scalar complex mass excitations lead to concrete predictions for GW spectra with power-law growth index $n_{GW}=0.7~-~1.2$ for the signals available for the future GW detectors. This could be an incredible possibility to test quantum gravity theory with gravitational wave signals, as well as with the other cosmologigal probes.} 
Most importantly, scalar complex mass excitations can induce a blue-tilted GW spectrum. When the tensor index $n_{GW}$ falls between $0.7$ to $1.2$, the GW signal will be detectable by future GW interferometers. This could present a novel and unforeseen chance to test infinite derivative quantum gravity theory in the arena of multi-messenger cosmology \footnote{See Ref.\cite{Addazi:2021xuf}
for a recent review on quantum gravity phenomenology in Multi-messenger physics.}.

%\section{Appendix}

%We can rewrite the relevant roots involved in the scalar field solution, with $m_{1}^2>0,m_{2}^2>0$
%as follows:
%$$\sqrt{9-4\Big(\frac{m_{1}^2}{H^2}\pm i\frac{m_{2}^2}{H^2}\Big)}$$
%\begin{equation}
%\label{jajkl}
%=\frac{1}{\sqrt{2}}\sqrt{ 9- \frac{4m_{1}^{2}}{H^2}+\sqrt{\Big( 9- \frac{4m_{1}^{2}}{H^2}\Big)^2+\frac{16m_{2}^4}{H^4}  } }\mp \frac{i}{\sqrt{2}}\sqrt{ -9+\frac{4m_{1}^{2}}{H^2}+\sqrt{\Big( 9- \frac{4m_{1}^{2}}{H^2}\Big)^2+\frac{16m_{2}^4}{H^4}  } }
%\end{equation}

\vspace{0.1cm
}

\noindent \textbf{Acknowledgements.}

This work is supported in part by the National Key Research and Development Program of China Grant No. 2021YFC2203004 (S.P.).
AA work is supported by 
National Science Foundation of China (NSFC) No.12350410358;
the Talent Scientific Research Program of
College of Physics, Sichuan University, Grant No.1082204112427 \&
the Fostering Program in Disciplines Possessing Novel Features for
Natural Science of Sichuan University, Grant No.2020SCUNL209 \& 1000
Talent program of Sichuan province 2021.  
S.P. is supported by Project No. 12047503 of the National Natural Science Foundation of China, by JSPS KAKENHI No. JP24K00624, and by the World Premier International Research Center Initiative (WPI Initiative), MEXT, Japan.
The work of AT was supported by the National Natural Science Foundation of China (NSFC) under Grant No. 12347103 and STFC grant ST/T000791/1.

	\bibliographystyle{utphys}
\bibliography{cosmo}

\end{document}